\RequirePackage[2020-02-02]{latexrelease}
\documentclass[%
aps,prl,amsmath,amssymb,
reprint,%
]{revtex4-2}

\usepackage{hyperref}
\usepackage{graphicx}% Include figure files
\usepackage{dcolumn}% Align table columns on decimal point
\usepackage{bm}% bold math
\usepackage{lipsum}
\usepackage{braket}
\usepackage[version=4]{mhchem}
\usepackage[utf8]{inputenc}
\usepackage[T1]{fontenc}
\usepackage{textgreek}
\usepackage{newtxtext}
\usepackage{newtxmath}

\usepackage{etoolbox}
\usepackage[capitalise]{cleveref}

\usepackage[dvipsnames]{xcolor}
\definecolor{correction}{HTML}{a51414}

\newcommand{\unit}[1]{{\,\mathrm{#1}}}
\newcommand{\mg}[1]{{\color{black}#1}}

\usepackage[toc,page]{appendix} % Added by Chulin

\newcommand{\sub}[1]{_\textrm{#1}} %Thomas: shortcut command to nonitalicize subscripts

\makeatletter
\def\@email#1#2{%
 \endgroup
 \patchcmd{\titleblock@produce}
  {\frontmatter@RRAPformat}
  {\frontmatter@RRAPformat{\produce@RRAP{*#1\href{mailto:#2}{#2}}}\frontmatter@RRAPformat}
  {}{}
}%
\makeatother
\begin{document}

%\preprint{AIP/123-QED}
\preprint{APS/123-QED}

\title[]{Direct electron cooling at millikelvin temperatures with quantum-well heat pump}
% Force line breaks with \\

% \thanks{These two authors contributed equally}
%  \altaffiliation{xxx}%
\author{Thomas Douglas}
\thanks{These two authors contributed equally}
\affiliation{%
Department of Electrical and Computer Engineering, Northwestern University, Evanston, IL 60208%\\This line break forced% with \\
}%  \email{Second.Author@institution.edu.}
\author{Chulin Wang}
\thanks{These two authors contributed equally}
\affiliation{%
Department of Electrical and Computer Engineering, Northwestern University, Evanston, IL 60208%\\This line break forced% with \\
}
% \affiliation{ 
% Authors' institution and/or address%\\This line break forced with \textbackslash\textbackslash
% }%
%\author{Lucia Steinke}
%\affiliation{Department of Physics, University of Florida, Gainesville, FL 32611
%\\This line break forced% with \\
%}
\author{Matthew A. Grayson$^*$}
\email{m-grayson@northwestern.edu.}
%  \homepage{http://www.Second.institution.edu/~Charlie.Author.}
\affiliation{%
Department of Electrical and Computer Engineering, Northwestern University, Evanston, IL 60208%\\This line break forced% with \\
}%

\date{\today}% It is always \today, today,

\begin{abstract}
Unconventional solid-state cooling is proposed whereby the electron subband degeneracy of a semiconductor quantum well is cycled at cryogenic temperatures. By %This solid-state electrostatic heat pump employs ``
``expanding'' electrons to higher subband degeneracy and ``compressesing'' to lower degeneracy in an Otto cycle, a stack of ultra-wide 150\,nm GaAs quantum wells filling a $1\unit{cm^3}$ volume could be operated as a heat pump to 
pump 26\,\textmu{W} at 100 mK. Operated as a refrigerator, it can %cool a factor of 2 in temperature from 150 mK to 75 mK or can 
piggy-back onto dilution refrigerators to cool from $T_\mathrm{base}$ to $T_\mathrm{base}/4$. %Such a device would be 
% the cooling power per unit device volume is estimated to reach $9\unit{mW/cm^3}$  with a hot-side temperature of $300$ mK,
Possible cooling applications include quantum computers and infrared detectors.
\end{abstract}

\maketitle

% \def\thefootnote{*}\footnotetext{These authors contributed equally to this work}\def\thefootnote{\arabic{footnote}}
% % text text text\footnote{normal footnote}

%\section{Introduction}

Sustainable and efficient millikelvin cryogenic cooling is essential for supporting cutting-edge applications such as solid-state quantum computation as well as other low-temperature experiments such as infrared detectors \cite{gardner2006james,agnese2014search} or for studying quantum states with long coherence times such as the quantum Hall effect \cite{levitin2022cooling}. 
\mg{In the ultralow temperature range %, defined below the limit of evaporative cooling from vacuum pumped homogeneous cryofluids $T < 300 \mathrm{mK}$ 
\cite{pobell2007}, only a handful of %laboratory-proven 
historical cryostat refrigeration mechanisms have proven viable: adiabatic electronic demagnetization of paramagnetic salts (1934) \cite{giauque1933}, adiabatic nuclear demagnetization of metals (1956) \cite{kurti1956}, Pomeranchuk cooling (1965) \cite{anufriev1965}, and He$^3$:He$^4$ dilution refrigeration (1965) \cite{das1965}.  %The rarity of useful refrigeration mechanisms in this ultralow range frames the need for new means of cryogenic refrigeration.
}  %To achieve millikelvin temperatures, t
The most common technology today is the \ce{^3He}/\ce{^4He} dilution refrigerator \cite{das1965realization}, which, however, requires \ce{^3He} - a scarce and non-renewable resource also needed for Pomeranchuk cooling. Demagnetization refrigeration %using polarizable salts 
demands significant overhead, requiring ancillary multi-Tesla superconducting electromagnets, an accompanying power-supply, and hard-to-shield stray fields \cite{figueroa2015searching,adams2020first}.
And in the end, all these methods are ineffective for cooling electrons \mg{to ultralow temperatures} since the systems that are directly cooled have vanishing coupling to the \mg{conduction} electrons at millikelvin temperatures. % decoupling from the system being directly cooled  between electron and phonon systems due to the vanishing electron-phonon coupling at lower temperatures. 
As a result, the base temperature of the electron system is frequently tens of millidegrees above that of the nominal bath base temperature.

In this Letter, a novel approach for sustained cooling is proposed that utilizes \mg{conduction} electrons, themselves, %directly 
as the refrigeration medium \cite{grayson2022cryogenic}.
This work proposes a device which expands the phase space of a two-dimensional electron system (2DES) from one to multiple subbands to generate cooling power, similar to the \ce{^3He} phase-space expansion utilized in dilution refrigerators. 
\mg{The efficiency of this subband degeneracy refrigerator (SDR) %The weak electron-phonon coupling  
increases at extreme low temperatures as the electron-phonon coupling vanishes, exhibiting} %becomes an \textit{advantage}, as it insulates the cooled electron system.
%Realizing a solid-state active cooling mechanism at the coldest end of the temperature spectrum will reduce the volume of %solid-state cryocoolers by orders of magnitude, increase hold times and improve thermal stability. 
%The loss in efficiency due to electron-phonon coupling decreases rapidly at lower temperatures, so this refrigeration method has 
no fundamental low-temperature limit.
%
%\revis{The applications of such a device are broad, as viable loads include any system in which good electrical contact can be made, including semiconductor, normal metal, and superconducting systems. This includes spin-tronic devices, such as  those composed of semiconductor heterostructures and Van der Waals materials. The application of this refrigerator to superconducting devices is limited by the low-entropy ground state which Cooper pairs enter well below the superconducting critical temperature, however, at temperatures closer to the superconducting critical temperature, electron cooling remain neccessary and highly beneficial.  }
Since coherence times typically increase with decreasing temperature \cite{duan1998reducing}, this may help to extend quantum coherence for solid-state and superconducting quantum information manipulation and storage.

%\section{Device Design}

\begin{figure}[!htbp]
    \centering
    \includegraphics[width=1.0\linewidth]{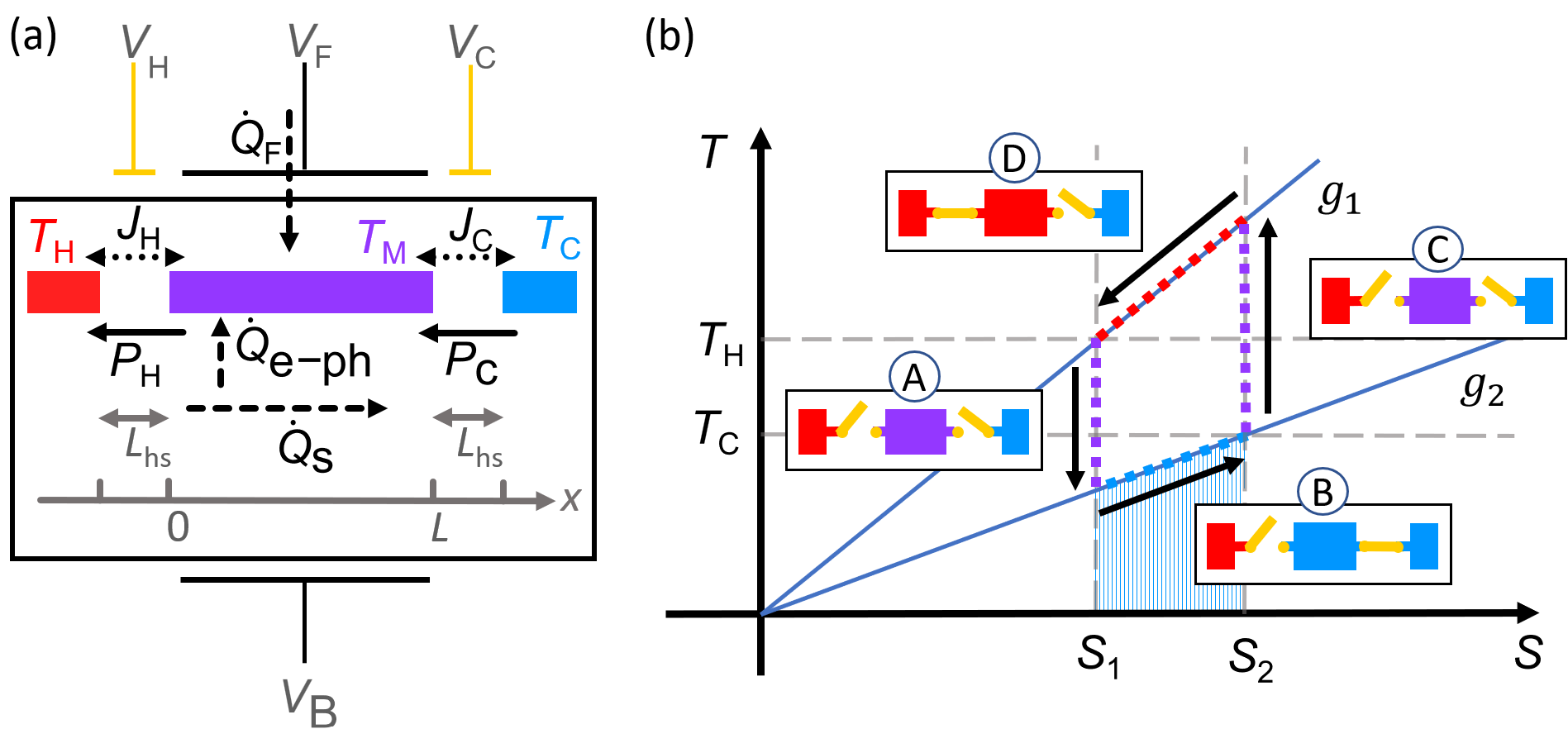}
    \caption{\mg{SDR} electron heat pump schematic and temperature-entropy diagram. (a) The central 2DES (purple) acts as working medium at variable temperature $T_\mathrm{M}$, electrostatically expanded/compressed into high/low subband degeneracy with the front and back piston gates $V_{\rm F}$ and $V_{\rm B}$. These interact with the heat sink (red) and cold load (blue) electrons, mediated by the heat-switch gates $V_{\rm H}$ and $V_{\rm C}$, with heat flows as defined in the body of the text.
    (b) Temperature-entropy diagram of the Otto refrigeration cycle: steps (A) and (C) show adiabatic expansion/compression, respectively, while (B) and (D) show isochoric heat transfer from cold load and to heat sink, respectively.\label{fig:device_thermalCircuit}
    }
\end{figure}

The \mg{SDR} electron heat pump is based on a semiconductor heterostructure with tunable quantum well (QW) subband occupancy.  This is achieved via metal gates on both front and back surfaces, following Rego and Kirczenow \cite{rego1999electrostatic} who first theoretically proposed a {\it single-shot} expansion of subband degeneracy $g$ of a two-dimensional electron system (2DES). However, their single-shot operation % is adiabatic, meaning no entropy is removed from the 2DES and the 
means any drop in electron temperature would be rapidly recovered through a heat leak from the phonon bath on the time-scale of the electron-phonon coupling, requiring unrealistically thin QW substrates. The present work extends this subband-expansion concept to a \textit{cyclic heat pump} operating according to a thermodynamic Otto cycle, achieving sustained cryogenic cooling power calculated to be effective for realistic substrate thicknesses, $t = 50$\,\textmu{}m. Just as a standard heat pump manipulates a working gas medium %of a liquid or molecular gas manipulated by 
with mechanical pistons and thermal switches, here a working medium of electrons in a QW is manipulated by electrostatic gates which serve as pistons and heat switches.

\Cref{fig:device_thermalCircuit}a shows a schematic of the proposed \mg{SDR} device. The central region (purple) of the QW 2DES is at variable temperature $T_{\rm M}$. The front and back gates (black) are biased to voltages $V_{\rm F}$ and $V_{\rm B}$, which are referred to as the piston gates for compression and expansion analogous to the pistons of a mechanical heat pump. These gates compress the central 2DES to low subband degeneracy $g_1$ or to expand the central 2DES to high subband degeneracy $g_2$.
The heat switch gates (gold) biased with voltages $V_\mathrm{H}$ and $V_\mathrm{C}$ can electrostatically deplete the electrons that connect the working medium to the cold load and heat sink, respectively, so that direct electronic conduction of heat across the entire device is inhibited throughout the refrigeration cycle. 
The various heat flows in \cref{fig:device_thermalCircuit}a are indicated with arrows, indicating the direction of heat flow under during Otto-cycle operation, where the solid black arrows indicate cooling power, dashed arrows represent Fourier heat leaks, and dotted arrows represent Joule heating from the depletion and repopulation of the 2DES under the heat-switch gates. Note that there is no Joule heating in the central region since no charge current flows there. Wires that connect to the gate terminals as well as the gate metals themselves will be made of superconductor, such as aluminum, whose thermal conductivity is negligible, eliminating the heat leak through those electrical contacts \cite{bardeen1959theory,satterthwaite1962thermal}. Normal metal wires, such as gold, will connect the device to the external heat sink and the load to-be-cooled.

%\section{Working Principle}

The cooling power of this \mg{SDR} device is generated by operating the working medium of electrons in an Otto cycle \cite{wu2003thermodynamic}.  \Cref{fig:device_thermalCircuit}b shows this $T$-$S$ diagram,  which is comprised of adiabatic expansion, isochoric heating, adiabatic compression, and isochoric cooling. The two $T(S)$ curves for degenerate electron systems with different $g$, can be obtained via Sommerfeld expansion \cite{Ashcroft} and are plotted in blue solid lines in \cref{fig:device_thermalCircuit}b,
\begin{equation}
    S = \frac{\pi m^* k_\text{B}^2 A\sub{QW}}{3 \hbar^2} gT_\mathrm{e} = KgT_\mathrm{e},
    \label{eq:entropy}
\end{equation}
where $k_\text{B}$, and $T\sub{e}$ are the Boltzmann constant and the electron temperature, and the proportionality factor $K$ is material-dependent through the electron effective mass $m^*$ and geometry dependent through the device area $A_\mathrm{QW}$.

The four legs of the Otto cycle in \cref{fig:device_thermalCircuit}b can be described as follows, with the inset reproductions of \cref{fig:device_thermalCircuit}a in miniature for each leg. (A) Both heat switches are disconnected. The temperature of the working medium drops from $T_{\rm H}$ to $(g_1/g_2)T_{\rm H}$ via an adiabatic expansion by biasing the front and back piston gates. (B) The right cold-side heat switch $V_\mathrm{C}$ is connected, coupling the working medium to the cold load at $T_{\rm C}$, and heat is drawn away from the cold load to the working medium in an isochoric heat flow. (C) Adiabatic compression is achieved by biasing the piston gates appropriately with both heat switches disconnected, and the temperature of the working medium is increased from $T_{\rm C}$ to $(g_2/g_1)T_{\rm C}$. (D) The left hot-side heat switch $V_\mathrm{H}$ is connected, coupling the working medium to the heat sink at $T_{\rm H}$, and heat is ejected from the working medium in an isochoric heat flow.

In the Otto cycle calculation below, we determine that MHz to GHz cycle frequencies can supply sufficient cooling power to compensate for parasitic heat loads, with excess cooling power to operate as a refrigerator. For a given operating frequency $f$ and heat sink temperature $T_{\rm H}$, the cooling power can be calculated from the shaded region of the $T$-$S$ diagram of \cref{fig:device_thermalCircuit}b,
\begin{equation}
    P_\textrm{c} = f \int_{S_1}^{S_2}{T_{g_2}(S)\, dS} = f \frac{K}{2 g_2}\left[(g_2T\sub{C})^2 - (g_1T\sub{H})^2\right].
    \label{eq:P_C}
\end{equation}
The primary source of electron heating %in the device that will counteract this cooling power 
is %direct 
electron-phonon coupling in the QW. It has been shown theoretically \cite{price1982hot} and experimentally \cite{MITTAL1996537} that the local heat exchange from phonons at $T_{\rm ph}$ to electrons at $T_{\rm e}$ follows a power law,
\begin{equation}
    \dot{Q}\sub{e-ph} = \left[3.3\times 10^6\tfrac{\unit{W}}{\unit{m^2K^5}}\right]\, \frac{A_{\rm QW}}{\sqrt{n}}\frac{1}{L}\int_0^{L} T\sub{ph}^5(x) - T\sub{e}^5\, dx,
    \label{eq:e-ph}
\end{equation}
where $n$ is the QW electron density. All quantities are in SI units. QW electrons are expected to have much faster thermal diffusion times than the phonons and are therefore modeled as having uniform temperature $T_\mathrm{e}$ across the width of the QW. The position-dependent phonon temperature is $T_{\rm ph} (x)$, where $x$ measures the length from the hot side at $x = 0$ to the cold side at $x = L$ (\cref{fig:device_thermalCircuit}a, bottom grey axis). The Fourier heat conduction in the lattice %density through the lattice is 
$\dot{Q}_\mathrm{S}$ is approximated as %being 
time independent under steady-state, %, %operation of the refrigerator and 
%modeled with one-dimensional Fourier heat flow. 
%The phonon temperature gradient will create a phonon "backflow" of heat from hot to cold reservoirs. This is 
characterized by the phonon thermal conductivity $\kappa_{\rm ph} = \frac{1}{3}  C_{\rm ph} v_{\rm ph} l_{\rm ph}$, where $v_{\rm ph}$ and $l_{\rm ph}$ are the temperature-independent phonon sound velocity and mean free path, respectively, and
$C_{\rm ph}(x) = 2 \pi^2 k_{\rm B}^4 T^3/(5 \hbar^3 v_{\rm ph}^3)$ is the phonon specific heat \cite{landau2013statistical}: % For the total heat flow, this gives
\begin{equation}
    \dot{Q}_{\rm S} = \frac{W t}{L_{\rm G} + 2L_{\rm hs}}\frac{\pi^2 k_{\rm B}^4 l_{\rm ph}}{30 \hbar^3 v_{\rm ph}^2}(T_{\rm H}^4 - T_{\rm C}^4),
\end{equation}
Here $W$, $t$, $L_{\rm hs}$ are the sample width, substrate thickness, and heat switch gate length, respectively. 

Since there is no charge flow, %during heat exchange 
the only source of Joule heating $J$ comes from electrons depleting and repopulating the area beneath the heat-switch gates: %, which can be reduced by including an oppositely biased reservoir gate immediately adjacent to the heat-switch gate of equal dimensions whose accumulated charge can accommodate the depleted charge under the heat switch. % TODO edit the heat storage reservoir look for citations
%With such a design, Joule heating $J$ is given by
\begin{equation}
    J = I^2R = 2f^2 \frac{ne}{\mu} L_\textrm{hs}^3 W,
\end{equation}
where $L_{\rm hs}$, $f$, $n$, $\mu$, and $e$ are the length of the heat switch gates, operating frequency, the QW electron density, mobility, and charge, respectively.

%The upper bound for h
High frequency operation will be limited by the thermal diffusion time $\tau_{\rm D} = L^2 / \alpha$ for electron heat to equilibrate along the QW length, %$L$ %during thermal equilibration of the central QW electrons with the cold load or heat sink. 
%This timescale is given by 
where $\alpha = \frac{2}{3} \frac{\mu E_\textrm{F}}{e}$ is the electron thermal diffusivity.   Since thermal equilibration must occur for both the hot and cold phases of each cycle, the cycling frequency becomes $f = 1 / 2\tau_{\rm D}$. The result is a frequency-dependent length constraint on the QW: %can be interpreted as a QW length constraint given a frequency-dependent thermal diffusion length: 
\begin{equation}\label{eq:L_of_f}
    L^2(f) = \frac{\alpha}{2f}.
\end{equation}
whereby the device length $L(f)$ represents an upper geometric bound for being able to operate at %the length of a device operating with given 
frequency $f$. % and provides maximum cooling power. 
Geometrically, one must then choose the lateral width $W(f) = A_{\rm QW} / L(f)$ according to this length constraint to hold the total device area $A_{\rm QW}$ constant.  Thus, a given device geometry will be optimized for a particular operation frequency $f$.

%\section{Material and Device Parameters}
\begin{figure}[h]
    \centering
    \includegraphics[width=1.05\linewidth,trim=0 0 0 0, clip]{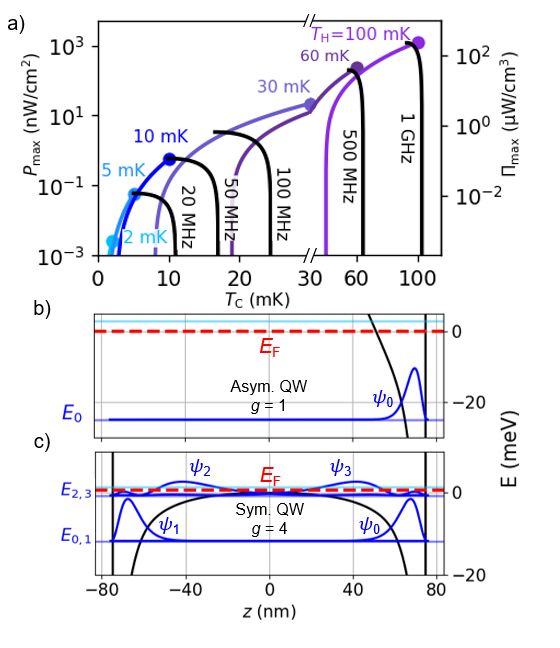}\\
    \caption{(a) \mg{SDR} calculated maximum %net 
    cooling power $P_\mathrm{max}$ %as a function of the 
    versus cold-side base temperature $T_\mathrm{C}$ % \leq T_\mathrm{H}$ 
    for various heat sink temperatures (colored curves) for  $T_\mathrm{H}=$ 2 to 100 mK.
    Colored dots represent %show maximum cooling power in 
    pure heat-pump mode  
    $T_\mathrm{C} =  T_\mathrm{H}$, and black curves indicate
    lines of constant cycling frequency $f$.
    Hartree calculations of (b) %the 
    asymmetric %imbalanced 
    and (c) symmetrically biased %balanced %condition for a 
    150 nm-wide GaAs QW %as 
    controlled by front- and back-piston gates. %For the i
    Imbalanced QW %only the ground state wavefunction 
    occupies %has 
    single-subband %occupancy 
    $g_1 = 1$ at energy $E_0$. %is occupied, resulting in a subband degeneracy of 
    %The b
    Balanced QW %, on the other hand, occupies %both  the 
    degenerate ground states $E_{0,1}$ and %first 
    excited states $E_{2,3}$ %degenerate subbands from 
    at opposite sides of QW with degeneracy $g_2 = 4$.}
    \label{fig:nW_cooling_power}
\end{figure}

A self-consistent Hartree calculation of a GaAs QW %is used as an example to 
demonstrates the viability of fourfold subband degeneracy in the electron heat-pump. \Cref{fig:nW_cooling_power}b,c show the balanced $g= 4$ and imbalanced $g = 1$ configurations of an ultra-wide QW $d_{\rm QW} = 150$ nm, with density $n= 10^{12}\ \rm cm^{-2}$ and effective mass  $m^* = 0.067 m_0$.

%\section{Cooling Power}
With the above model, the net cooling power $P_\mathrm{net}$ of an Otto cycle refrigerator with degeneracy ratio $g_\mathrm{2}$:$g_\mathrm{1}$ = 4:1, optimal device length $L(f)$ from \cref{eq:L_of_f}, and heat switch length $L_{\rm hs} = 1$\,\textmu{m} can be estimated, considering the heat loss terms defined earlier:
\begin{equation}
    P_{\rm net} = P_{\rm C} - J - \dot{Q}_\textrm{e-ph} - \dot{Q}_{\rm S}.
\end{equation}
For $T_{\rm H} = 2\ \rm mK$ to $100\ \rm mK$, \cref{fig:nW_cooling_power}a shows the calculated maximum net cooling power $P_{\rm max}$ versus cold-side temperature $T_\mathrm{C}$, with optimal frequency of operation $f$ (black lines). 
Simulations assume %a QW with electron density $n = 1 \times 10^{12}\unit{cm^{-2}}$, 
electron mobility $\mu = 2.0 \times 10^7\unit{ cm^2/Vs}$ \cite{reichl2014increasing,banerjee2018observation,hwang2008limit,manfra2014molecular} known to be achievable below $1\ \rm K$ \cite{schlom2010upward}, and GaAs electron effective mass $m^* = 0.067m_{\rm 0}$. 
Calculations assume a substrate thickness of $t = 50$\,\textmu{}m, achievable by mechanically or chemically thinning the sample, with even thinner micron-scale substrates achievable with remote epitaxial growth \cite{kim2017remote} to further minimize substrate phonon back-flow.
The substrate phonon velocity is $v_{\rm ph} = 4000\unit{m/s}$, and the phonon mean-free path $l_\mathrm{ph} = 20$\,\textmu{m} is estimated from the lowest experimentally measured bulk GaAs thermal conductivity in $p$-type Zn-doped GaAs substrates ($N_\mathrm{Zn} = 1.9 \times 10^{19}/\mathrm{cm}^3$) \cite{carlson1965GaAs-Zn}.

Whereas the cooling powers above were defined per-unit-area for a single QW, one can %enhance cooling power density further with 
consider a volumetric design of stacked QWs.  Such a vertical stack of devices operating in parallel, each of thickness $t = 50$\,\textmu{m}, can realize %a multi-heat-pump design that is connected both electrically and thermally in parallel. 
volumetric cooling power per unit of device volume, $\Pi_\mathrm{max} = P_\mathrm{max}/t$, simply by
dividing the  power per unit area in \cref{fig:nW_cooling_power}a by the substrate thickness $t$ % yields such a .
(right vertical axis).  
% of \cref{fig:nW_cooling_power}a estimates volumetric cooling per cm$^3$ under such a design.

\mg{The %subband degeneracy refrigerator (SDR) 
SDR should be able to directly cool electrons in any electrically connected sample, from monolithic quantum wells and Van der Waals materials to semiconductor quantum dots and superconducting qubits. 
Similarly, it should be able to pump excess heat from a transmission line %to limit parasitic heat flow on 
for signal and drive lines of a quantum computer.  All these applications rely on the fact that the cold side is a virtual open-circuit relative to the hot side: for each phase of the cooling cycle, one or more of the heat-switch gates are open-circuited, so there is never a direct electrical conduction path from the cold side to the hot side.  This allows the cold side to be fixed to the electrical ground of the %cold load 
element-to-be-cooled and the hot side to serve as an electrically floating thermal sink.}

\mg{As a refrigeration device, the SDR can be used in multiple ways: purely as a heat pump to maximize heat flow $P_{max}$ with no temperature differential $T_\mathrm{C} = T_\mathrm{H}$; as a cooler to achieve maximum temperature difference $T_\mathrm{H} - T_\mathrm{C}$ with no cold-side cooling power $P_\mathrm{c} = 0$; or at intermediate temperatures to pump non-zero heat $P_\mathrm{c}$ actively from the cold side.} % minimize the cold side temperature with no net heat 
The first mode of SDR operation is as an electron heat-pump with no thermal differential. Here, $T_\mathrm{C} = T_\mathrm{H}$ as indicated with colored circles in \cref{fig:nW_cooling_power}a, and heat is pumped directly out of the electrons at the cold side and expelled to the heat sink at the same temperature. 
\mg{
This is relevant anywhere there is electron heating above the local ambient temperature. This occurs in superconducting transmission lines where thermal leakage excites quasiparticles and}
%This is particularly relevant 
at the lowest cryogenic temperatures where reduced electron-phonon coupling leads to electron temperatures much higher than the surrounding cooled phonon system. The results %in \cref{fig:nW_cooling_power}a 
show that in heat-pump mode, %(colored circles), 
this single stage cooler can provide net heat pumping power-per-unit-area (volume) of $P_{\rm net} = 1.3$\,\textmu{W/cm$^2$} ($\Pi_{\rm net} = 26$\,\textmu{W/cm$^3$}) at $100\unit{mK}$, as well as $P_{\rm net} =24$\, nW/cm$^2$ ($\Pi_{\rm net} = 480$\,\textmu{W/cm$^3$}) at $30\ \rm mK$.  %In the stacked volumetric cooler, one can achieve an ambitious  of cooling power-per-unit-volume in heat-pump mode at $T_\mathrm{H} = T_\mathrm{C} = 100\unit{mK}$. 
\mg{For removing excess heat from transmission lines %cooling 
with 
the 
SDR %should be operated 
in heat-pump mode, 
%to actively cool large heat loads to the ambient local temperature. In this case, 
the SDR cold-side could be thermally shorted to the ground shield of the superconducting transmission line using a normal metal lead. %, thereby cooling any super-thermal quasielectron and quasihole excitations of the superconductor to local ambient temperature. 
} 

The second mode of SDR operation is an electron cooler, %refrigerator, 
where the cold load is cooler than the heat sink $T_\mathrm{C}<T_\mathrm{H}$.  Fourier back-flow of heat through the substrate phonons makes refrigeration less effective at the highest temperatures. But as the $T^3$ specific heat of the phonon bath is further suppressed at lower temperatures, performance approaches the ideal limit, with a reduction in electron temperature approaching the degeneracy ratio $g_2/g_1 = 4$, such as with $T_\mathrm{H} = 30\unit{mK}$ where the device can cool down to $T_\mathrm{C} = 8\unit{mK}$, as seen in \cref{fig:nW_cooling_power}a.
%The frequency of operation can also be read off of \cref{fig:nW_cooling_power}a.
\mg{
When refrigerating a sample to lower temperatures with the SDR in cooler mode, the SDR cold-side can be thermally shorted with a normal metal wire to the sample at its ground pin.  The ground pin is, in turn, connected to the sample’s external circuit ground through a thermally insulating but electrically conducting superconducting lead.}

Recall that the black curves trace the optimal frequency of performance. % of a single device operating at its optimal frequency across a set of $T_\mathrm{H}$ operation curves.
%As previously described, choosing $f$ determines an optimal $L_q$ due to  Eq.\,\eqref{eq:L_of_f}. The intersection of the black, constant frequency and colored, constant $T_\mathrm{H}$ curves can 
These lines give an impression of typical frequencies needed for operation at a given $T_\mathrm{H}$ and $T_\mathrm{C}$. For example, in heat-pump mode with $T_\mathrm{H}=T_\mathrm{C}=60$\,mK, the operating frequency is $f = 500$\,MHz, and in refrigeration mode with the same hot-side temperature, base temperature of $T_\mathrm{C} = 20$\,mK is achieved with just over $f = 50$\,MHz. 
\mg{For best performance, it is important that the sample of interest also be thermally isolated.  At low enough temperatures, given that the lattice thermal conductivity and electron-phonon coupling vanish, and the Kapitza resistance \cite{pollack1969kapitza} at material interfaces also increases, most scenarios for mechanical sample mounting achieve the necessary thermal isolation. More important is the thermal isolation of the electrical leads, which can be achieved by using superconducting leads which conduct electrically but have low thermal conduction.}

\mg{
%To achieve this, the mixing chamber of the dilution refrigerator could be loaded with a silver sintered heat exchanger, which is in tu
%As an example experimental implementation, 
This SDR device 
can be attached to a dilution refrigerator to %applied to 
%could be implemented as follows to 
further reduce the base temperature. %enhance the cooling power 
%of a dilution refrigerator. %, for example. % or achieve lower base temperatures. %, for example. 
% this SDR can be piggy-backed onto a dilution refrigerator to achieve lower base temperatures. 
The mixing chamber of the dilution refrigerator could be loaded with a sintered silver heat exchanger and thermally sunk to the hot-side of the SDR with a normal metal wire, and then the cold side of the SDR could be connected to the sample.} %, with the sample under study coupled to the cold side.
To %eventually provide an all-solid-state alternative to %compete with 
further enhance %
%supplant dilution 
refrigeration, %with an all-solid , 
cascaded multistage designs can be considered to allow for higher hot-side temperatures while achieving greater thermal differentials to colder base temperatures. The lateral design of the subband degeneracy refrigerator could be particularly well-suited for cascading, allowing for simple, monolithic, planar cascade designs where the heat sink of the former stage serves as the cold load of the latter stage. 

Alternatively, the subband degeneracy refrigerator could be paired with other solid state cooling mechanisms
such as superconducting-insulating-normal metal junction cooling \cite{leivo1996efficient,o2012measurement} to achieve a composite cascaded all-solid-state alternative to dilution refrigeration. %\revis{, again with only the requirement of good electrical contact between series-connected refrigerators}.  
Other thermodynamic heat pump cycles besides the Otto cycle may lead to improved cooling efficiency or increased cooling power at higher temperatures where electron-phonon coupling is enhanced. This proposed mode of refrigeration therefore represents a novel physical mechanism for low-temperature solid-state refrigeration.

\begin{acknowledgments}
This material is based upon work supported by the U.S. Department of Energy, Office of Science, National Quantum Information Science Research Centers, Superconducting Quantum Materials and Systems Center (SQMS) under the contract No. DE-AC02-07CH11359. T.D. thanks the support of URAP funding from the Office of Undergraduate Research, Northwestern University. The authors also thank Lucia Steinke (U. Florida) for insightful conversations.

\end{acknowledgments}

\bibliography{solid_state_cooling}
% Produces the bibliography via BibTeX.
%%
\end{document}